\documentclass[runningheads]{llncs}

\usepackage[T1]{fontenc}
\usepackage{graphicx}
\usepackage{booktabs}
\usepackage{multirow}
\usepackage{url}
\usepackage{amsmath}
\usepackage[hidelinks]{hyperref}

\begin{document}

% ============================================================
% TITLE
% ============================================================

\title{Reddit Community Interventions and Cross-Platform Response: Migration and Policy-Response Discourse on Voat}

\titlerunning{Community Bans and Cross-Platform Response}

% ============================================================
% AUTHORS
% ============================================================

\author{Shahan Ahmed}

\authorrunning{S. Ahmed}

\institute{
Montclair State University\\
\email{ahmeds48@montclair.edu}
}

\maketitle

% ============================================================
% ABSTRACT
% ============================================================

\begin{abstract}

Platform moderation can disrupt online communities, but the responses
that follow do not necessarily take the form of direct migration.
Communities may relocate, reorganize, discuss moderation decisions,
express censorship concerns, or continue activity on alternative
platforms without clear evidence of user-level movement.

This study examines Reddit community bans and quarantines and their
relationship with activity and migration-related discourse on Voat.
Using data from the Multi-Platform Aggregated Dataset of Online
Communities (MADOC), we analyze interventions affecting
FatPeopleHate, GreatAwakening, MillionDollarExtreme, and
CringeAnarchy. We combine computational discourse screening with a
stratified independent human-validation procedure and population-adjusted estimates across 90-day pre- and post-intervention windows.

The results reveal heterogeneous cross-platform responses.
FatPeopleHate experienced a substantial increase in overall Voat
activity while the relative prevalence of policy-response and broader
relevant discourse declined. CringeAnarchy showed declines in direct
migration and policy-response discourse following its permanent ban,
although some estimates were sensitive to uncertain human labels.
GreatAwakening instead exhibited a modest increase in policy-response
discourse without clear evidence of increased direct migration.

These findings suggest that platform interventions do not produce a
uniform migration process. Overall destination-platform activity and
migration-related discourse may move in different directions,
highlighting the importance of distinguishing behavioral migration
from broader discussion of moderation, censorship, governance, and
community rebuilding.


\end{abstract}

% ============================================================
% 1. INTRODUCTION
% ============================================================

\section{Introduction}

Online platforms increasingly use community-level moderation to address
harassment, hate speech, misinformation, and other harmful behavior.
Interventions range from visibility restrictions and quarantines to
permanent community bans. Prior research shows that these actions can
reduce harmful activity on the source platform
\cite{chandrasekharan2017ban,chandrasekharan2022quarantined}.
Moderation, however, may also prompt users and communities to relocate,
fragment, rebuild elsewhere, or discuss censorship and platform
governance on alternative platforms
\cite{ribeiro2021migrations,monti2023resilient,buntain2023crossplatform}.

This creates an important measurement problem: \emph{discussion about
migration is not equivalent to migration itself}. References to Reddit,
Voat, bans, censorship, rebuilding, or relocation may indicate a
collective response to moderation without demonstrating that identifiable
users actually moved between platforms. Likewise, destination-platform
activity can increase sharply even while migration- or governance-related
discourse becomes a smaller share of the conversation. Distinguishing
overall activity from different forms of intervention-related discourse
is therefore necessary for interpreting cross-platform responses.

We examine Reddit interventions involving FatPeopleHate,
GreatAwakening, MillionDollarExtreme, and CringeAnarchy and their
corresponding activity and discourse on Voat. Using the Multi-Platform
Aggregated Dataset of Online Communities (MADOC) \cite{madoc2025}, we
analyze permanent bans as well as the quarantine and later permanent ban
of CringeAnarchy.

Our central research question is:

\begin{quote}
\emph{How do community bans and quarantines correspond with direct
migration discourse, policy-response discourse, and broader
intervention-related activity on an alternative platform?}
\end{quote}

We combine event-centered activity analysis with computational text
screening and independent human validation. For comparable cases, we
examine 90-day pre- and post-intervention windows. Automated indicators
screen for references to migration, Reddit and Voat, intervention,
censorship or free speech, community rebuilding, and governance
backlash. Because narrow migration patterns can misclassify general
platform discussion, we validate the measures using a stratified
manually coded sample and population-adjusted weights.

The study contributes in three ways. First, it separates explicit
direct-migration discourse from broader policy and platform responses.
Second, it compares multiple interventions using a common event-window
framework while preserving cross-community heterogeneity. Third, it
combines scalable computational screening with weighted human validation
rather than treating keyword matches as definitive evidence of
cross-platform movement.

The results reveal no uniform post-intervention trajectory.
FatPeopleHate shows a sharp increase in overall Voat activity while the
prevalence of policy-response and broader relevant discourse declines.
CringeAnarchy shows post-ban declines in direct-migration and
policy-response discourse, although some estimates are sensitive to
uncertain labels. GreatAwakening instead shows a modest increase in
policy-response discourse without clear evidence of increased direct
migration. Overall activity and migration-related discourse therefore
need not move in the same direction.

% ============================================================
% 2. RELATED WORK
% ============================================================
\section{Related Work}

Community-level moderation can reduce problematic behavior on the
platform imposing the intervention, although effects vary across
communities \cite{thomas2021behavior,cima2024greatban}.
Chandrasekharan et al. found that Reddit's 2015 bans of
FatPeopleHate and CoonTown were followed by user departures and large
reductions in hate-speech use among users who remained
\cite{chandrasekharan2017ban}. Across 15 banned subreddits, Trujillo
et al. similarly found heterogeneous changes in activity and
community-specific language \cite{trujillo2021echo}. Interventions
short of removal can produce different effects: Reddit quarantines
substantially reduced recruitment into targeted communities while
showing limited changes in the behavior of existing members
\cite{chandrasekharan2022quarantined}.

A related literature examines responses beyond the source platform.
Horta Ribeiro et al. studied The\_Donald and Incels after migration
from Reddit to dedicated websites and found declines in participation
alongside evidence that some content characteristics could become more
extreme \cite{ribeiro2021migrations}. Monti et al. directly examined
FatPeopleHate and GreatAwakening migration from Reddit to Voat,
finding that both partially migrated but that GreatAwakening was more
resilient and more likely to reconstruct prior social connections
\cite{monti2023resilient}. Migration can also continue across
successive platforms, as illustrated by QAnon-related movement from
Reddit to Voat and later from Voat to Poal
\cite{papasavva2024waiting}. Cross-platform responses following the
January 6 deplatforming likewise differed across alternative platforms
in attention, engagement, and discourse
\cite{buntain2023crossplatform}.

Together, these studies show that moderation can correspond with
departure, relocation, fragmentation, or reorganization rather than a
single uniform response. Our study complements user-matching approaches
by distinguishing what can be inferred from discourse itself. We
separate explicit direct-migration discourse from broader
policy-response discourse---including intervention, censorship,
governance, and rebuilding discussion---and from general relevant
platform discourse. This prevents references to an alternative
platform from being treated automatically as evidence that identifiable
users migrated.

% ============================================================
% 3. DATA AND METHODS
% ============================================================

\section{Data and Methods}

\subsection{Data, Event Windows, and Discourse Measures}

We use the Multi-Platform Aggregated Dataset of Online Communities
(MADOC), which provides harmonized records from multiple online
platforms, including Reddit and Voat \cite{madoc2025}. The analysis
focuses on Voat communities associated with Reddit communities that
experienced a major platform intervention.

Five intervention events are considered: the permanent ban of
FatPeopleHate on June 10, 2015; the permanent ban of
MillionDollarExtreme on September 10, 2018; the permanent ban of
GreatAwakening on September 12, 2018; the quarantine of
CringeAnarchy on September 27, 2018; and the permanent ban of
CringeAnarchy on April 25, 2019. The CringeAnarchy quarantine and
permanent ban are treated as distinct interventions.

Table~\ref{tab:interventions} summarizes the analytical windows.
FatPeopleHate, GreatAwakening, and the permanent CringeAnarchy ban
contain sufficient observations for pre--post comparison.
MillionDollarExtreme has no observations in the corresponding
pre-intervention window and is therefore described only after the
intervention. The CringeAnarchy quarantine has only five
pre-intervention interactions and is treated as descriptive rather
than as a primary comparative case.
\begin{table}[t]
\caption{Intervention cases and analytical sample sizes. Counts refer
to Voat interactions within the corresponding 90-day event windows.}
\label{tab:interventions}
\centering
\small
\begin{tabular}{@{}lclrrl@{}}
\toprule
Intervention & Date & Type & Pre & Post & Role \\
\midrule
FatPeopleHate
& 2015-06-10 & Ban
& 180 & 46,750 & Primary \\

MillionDollarExtreme
& 2018-09-10 & Ban
& 0 & 6,922 & Post only \\

GreatAwakening
& 2018-09-12 & Ban
& 3,727 & 36,051 & Primary \\

CringeAnarchy quarantine
& 2018-09-27 & Quarantine
& 5 & 30 & Descriptive \\

CringeAnarchy ban
& 2019-04-25 & Ban
& 369 & 2,048 & Primary \\
\bottomrule
\end{tabular}
\end{table}

For each intervention, we construct an event-centered window using the
official intervention date as day zero. The pre-intervention period
covers the 90 calendar days immediately preceding the intervention
(days $-90$ through $-1$), while the post-intervention period covers
the intervention date and the following 89 days (days $0$ through
$+89$). This produces comparable 90-day windows while preventing the
intervention date itself from being included in the pre-period.

The analysis is observational. Pre--post differences are interpreted
as temporal changes associated with an intervention rather than causal
effects. Other contemporaneous events may also influence activity and
discourse within the analyzed communities.

We use computational text screening to identify several forms of
intervention-related discourse. The screening procedure searches the
text of Voat interactions for references to Reddit, Voat, platform
interventions, censorship and free speech, migration or relocation,
community rebuilding, and governance backlash.

These indicators are combined into three nested analytical outcomes.
\emph{Direct migration} represents the narrowest category and requires
language that explicitly or strongly implies movement between
platforms, such as statements about moving from Reddit to Voat,
Reddit refugees, or migration following a ban. \emph{Policy-response
discourse} is broader and includes direct migration as well as
discussion connecting Reddit or Voat with moderation interventions,
censorship, free speech, rebuilding, or related governance responses.
Finally, \emph{relevant discourse} is the broadest measure and includes
any substantive reference captured by the intervention-related
screening system.

The three outcomes are intentionally distinguished because a reference
to Reddit, a ban, or censorship does not by itself demonstrate
cross-platform migration. Automated screening is therefore used
primarily to identify candidate discourse rather than to establish
behavioral movement of identifiable users.

\subsection{Manual Validation and Population Weighting}

Because intervention-related discourse is relatively rare and automated
keyword patterns can generate both false positives and false negatives,
we use stratified human validation. Candidate interactions were divided
within each event-period combination into four mutually exclusive
sampling strata: (1) strict migration matches, (2) broader
policy-response matches that were not strict migration matches,
(3) other relevant-discourse matches, and (4) interactions with no
detected signal.

Up to 25 interactions were sampled from each available stratum using a
fixed random seed, producing 609 manually coded records. A single
independent human coder evaluated each interaction for overall
relevance, direct migration, and a primary discourse category. The
primary categories were direct cross-platform migration, intervention
discussion, censorship or free speech, community rebuilding,
governance backlash, platform reference only, unrelated, and
ambiguous. Direct migration required an explicit or clearly implied
movement between platforms. Uncertain judgments were recorded as
\texttt{U} rather than being forced into a binary category.

Because the validation design deliberately oversampled interactions
containing automated signals, raw proportions in the validation sample
do not represent population prevalence. Each manually coded
observation was therefore assigned an inverse sampling-fraction weight

\[
w_h = \frac{N_h}{n_h},
\]

where $N_h$ is the number of population interactions in an
event-period sampling stratum and $n_h$ is the number manually sampled
from that stratum. Weighted estimates reconstruct event-period
population rates for direct migration, policy-response discourse, and
relevant discourse.

For the three intervention cases with adequate pre- and post-period
data, we report the difference in weighted prevalence as

\[
\Delta = p_{\mathrm{post}} - p_{\mathrm{pre}},
\]

expressed in percentage points. Ninety-five percent confidence
intervals account for the stratified validation design and finite
population corrections. These intervals quantify uncertainty arising
from the manual-validation sampling design; they should not be
interpreted as causal confidence intervals.

The primary analysis excludes uncertain human judgments from binary
classification. As a robustness check, we also calculate extreme
bounds in which all uncertain observations are alternatively treated
as negative or positive. This sensitivity analysis is used to identify
findings that depend heavily on the small number of ambiguous
observations, particularly when those observations carry large
sampling weights.

% 90-day pre/post design and analytical cases.

% Automated discourse definitions.

% Human validation, stratified sampling, weighting, confidence intervals.

% ============================================================
% 4. RESULTS
% ============================================================

\section{Results}

\subsection{Validation of the Discourse Measures}

Manual validation showed substantial differences across the three
computational discourse measures. After accounting for the stratified
sampling design, the broad relevant-discourse detector achieved high
recall ($0.974$) with an F1 score of $0.704$. The policy-response
detector was substantially more conservative, with precision of
$0.593$, recall of $0.378$, and an F1 score of $0.462$.

The strict direct-migration detector performed poorly as a standalone
classifier. Its weighted precision was $0.201$, recall was $0.130$,
and F1 was $0.158$. Inspection of disagreements showed that strict
patterns missed contextually expressed migration while also
occasionally classifying intervention or rebuilding language as direct
movement. Consequently, automated direct-migration matches are not
treated as definitive evidence of migration. The substantive
pre--post estimates below instead rely on the weighted human-coded
classifications.

These results indicate that computational screening is useful for
retrieving broad intervention-related discourse, but increasingly
narrow semantic claims require human validation.

\subsection{Cross-Event Pre--Post Changes}

\begin{figure}[htbp]
\centering
\includegraphics[width=0.88\textwidth]
{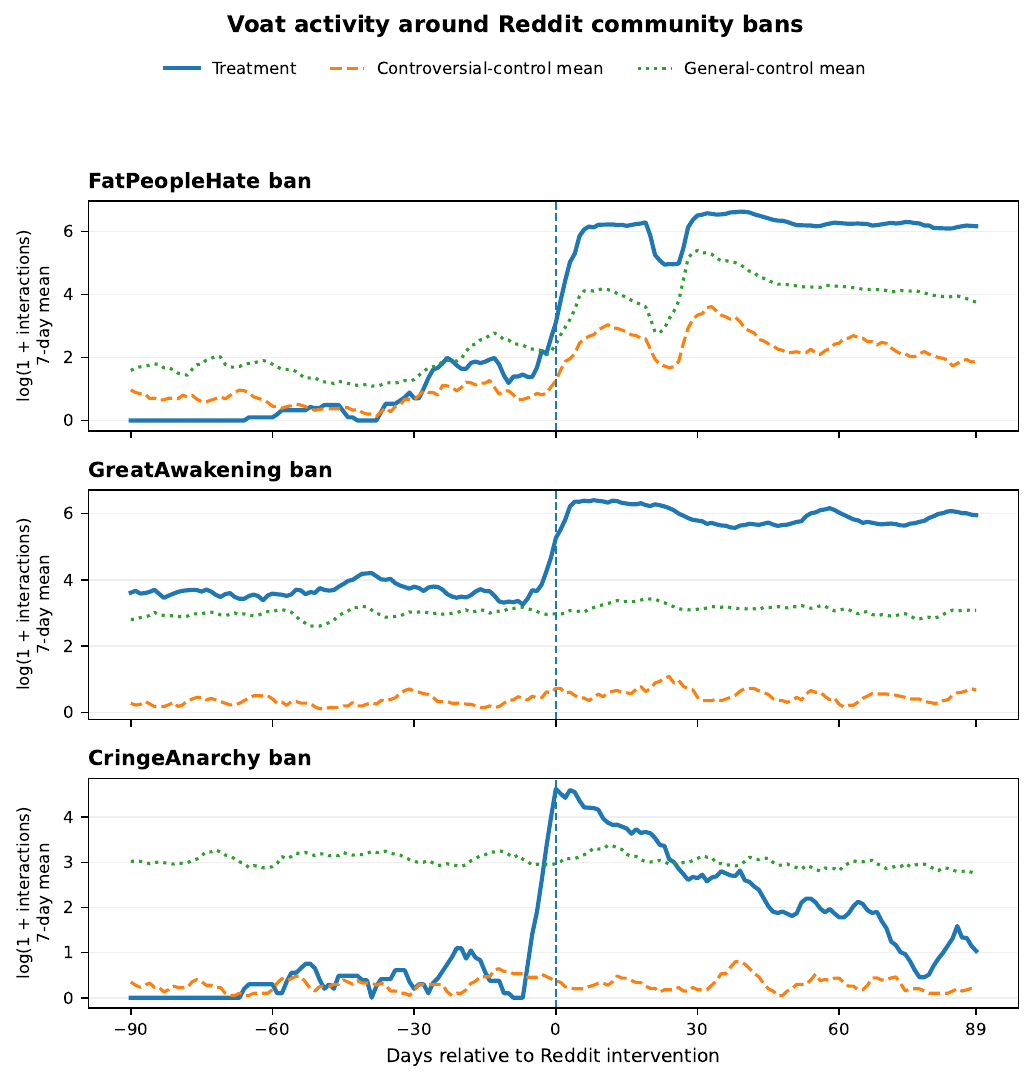}
\caption{Voat activity around three Reddit community bans. Lines show
7-day moving averages of $\log(1+\text{daily interactions})$; the
vertical dashed line marks the intervention date.}
\label{fig:activity}
\end{figure}

\begin{figure}[htbp]
\centering
\includegraphics[width=0.78\textwidth]
{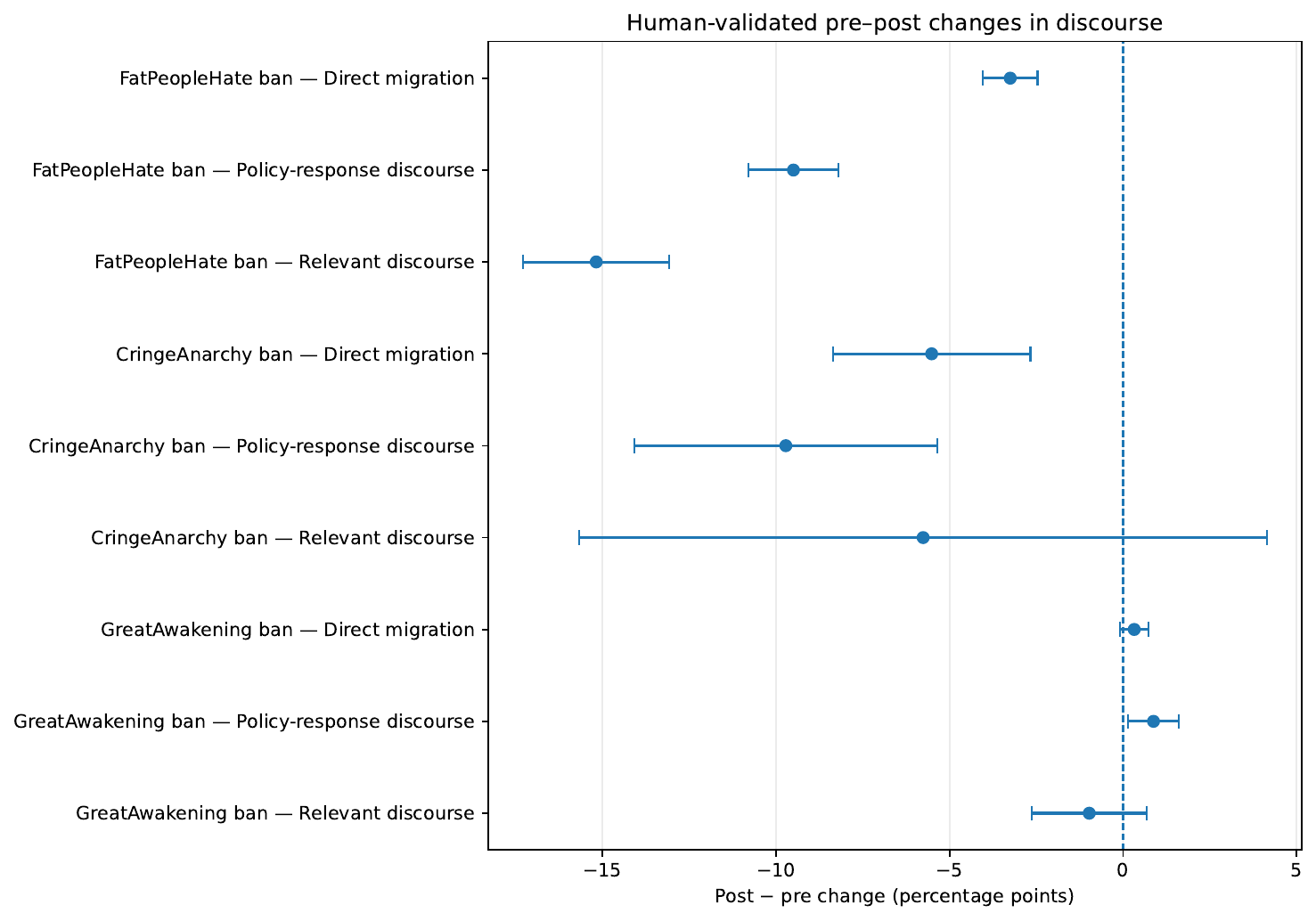}
\caption{Human-validated pre--post discourse changes. Points show
weighted percentage-point changes with 95\% design-based confidence
intervals.}
\label{fig:effects}
\end{figure}

Figure~\ref{fig:activity} shows overall Voat activity trajectories
around the three comparable permanent bans, while
Figure~\ref{fig:effects} summarizes the weighted human-validated
percentage-point changes and confidence intervals. Together, these
measures distinguish changes in overall activity volume from changes
in the composition of migration- and policy-related discourse.

FatPeopleHate displayed the clearest reduction in intervention-related
discourse. Direct-migration discourse decreased from $3.81\%$ before
the ban to $0.56\%$ afterward, a change of $-3.25$ percentage points
(pp; 95\% CI $[-4.04,-2.46]$). The decrease was considerably larger
for policy-response discourse, from $10.99\%$ to $1.50\%$
($\Delta=-9.50$ pp, 95\% CI $[-10.79,-8.21]$), and for broader
relevant discourse, from $17.80\%$ to $2.62\%$
($\Delta=-15.18$ pp, 95\% CI $[-17.29,-13.07]$).
The decreases in policy-response and relevant discourse remained
directionally robust under the extreme treatment of uncertain human
labels. The direct-migration estimate was more sensitive to those
uncertain observations and is therefore interpreted more cautiously.

CringeAnarchy showed a broadly similar directional pattern after its
permanent ban. Direct-migration discourse declined from $6.33\%$ to
$0.81\%$ ($\Delta=-5.52$ pp, 95\% CI $[-8.36,-2.67]$), while
policy-response discourse declined from $13.97\%$ to $4.25\%$
($\Delta=-9.72$ pp, 95\% CI $[-14.08,-5.36]$). Broader relevant
discourse also declined, but its confidence interval crossed zero
($\Delta=-5.76$ pp, 95\% CI $[-15.67,4.15]$). Moreover, the
CringeAnarchy estimates were sensitive to alternative assignments of
uncertain human labels. They therefore provide evidence of a
post-intervention decline in direct-migration and policy-response
discourse, but with weaker robustness than the principal
FatPeopleHate results.

GreatAwakening followed a different trajectory. Policy-response
discourse increased from $0.43\%$ to $1.30\%$, corresponding to a
$0.88$ pp increase (95\% CI $[0.14,1.61]$). Direct-migration
discourse increased only from $0.11\%$ to $0.43\%$, with a confidence
interval spanning zero ($\Delta=0.32$ pp, 95\% CI $[-0.08,0.73]$).
Relevant discourse decreased from $3.57\%$ to $2.59\%$, but this
difference was also uncertain ($\Delta=-0.97$ pp, 95\% CI
$[-2.64,0.69]$). The positive policy-response estimate was sensitive
to extreme treatment of uncertain labels and should therefore be
interpreted as suggestive rather than definitive evidence of an
increase.

\subsection{Small and Incomplete Intervention Cases}

The CringeAnarchy quarantine contained only five interactions in the
90-day pre-intervention window and 30 interactions in the post-window.
Although weighted estimates can be calculated for these records, the
extremely small population makes conventional comparative
interpretation inappropriate. We therefore retain the quarantine only
as descriptive evidence.

MillionDollarExtreme presents the opposite limitation: 6,922
interactions are available after the September 2018 ban, but no
observations are available in the corresponding 90-day pre-period.
Within the post-period, the weighted human estimates were $1.57\%$
for direct migration, $3.26\%$ for policy-response discourse, and
$5.69\%$ for relevant discourse. These values characterize
post-intervention discourse but cannot establish a pre--post change.

% ============================================================
% 5. DISCUSSION
% ============================================================

% ============================================================
% 5. DISCUSSION
% ============================================================

\section{Discussion}

The cross-case results show that Reddit interventions were not followed
by a uniform pattern of activity or migration-related discourse on
Voat. Instead, the cases exhibit different combinations of destination
activity, direct-migration discourse, and broader policy response. This
heterogeneity is consistent with prior work showing that community
responses to moderation depend on the characteristics of the affected
community and its members
\cite{trujillo2021echo,monti2023resilient}.

FatPeopleHate provides the clearest illustration of why activity volume
and discourse composition should be distinguished. Overall Voat
activity increased sharply around the Reddit ban
(Fig.~\ref{fig:activity}), yet the weighted prevalence of
policy-response and broader relevant discourse declined substantially.
The result therefore does not indicate an overall contraction of the
destination community. Rather, intervention- and migration-related
discussion became a smaller share of a rapidly expanding stream of
activity. A post-ban activity surge consequently should not be treated
as equivalent to a sustained increase in migration discourse.

CringeAnarchy followed a different temporal pattern. Activity began
increasing immediately before the permanent ban, peaked near the
intervention, and subsequently declined. This pre-ban mobilization
reinforces the need for caution in attributing the observed trajectory
solely to the intervention. Human-validated direct-migration and
policy-response discourse nevertheless declined between the 90-day
windows, although these estimates were less robust to uncertain labels
than the strongest FatPeopleHate findings.

GreatAwakening provides a contrasting case. Treatment-community
activity increased substantially around the ban and remained elevated,
while policy-response discourse showed a modest increase. Direct
migration, however, did not show a clearly distinguishable increase.
Thus, heightened discussion of platform governance or intervention
does not necessarily provide evidence of proportional increases in
direct cross-platform movement.

More generally, these results support separating at least three
dimensions of cross-platform response: overall destination activity,
explicit migration discourse, and broader policy or governance
discussion. Automated screening was effective for retrieving broad
candidate discourse but performed poorly for the narrow direct-migration
category. Human validation is therefore particularly important when
textual evidence is used to make claims about migration rather than
general platform response.

% ============================================================
% 6. LIMITATIONS AND CONCLUSION
% ============================================================

\section{Limitations and Conclusion}

Several limitations qualify the findings. The analysis identifies
migration-related discourse rather than matched cross-platform
identities, so it cannot establish individual-level migration, and
Voat is only one possible destination. The observational pre--post
design may reflect concurrent events or existing trends, particularly
for CringeAnarchy. Coverage is uneven: the quarantine is sparse,
MillionDollarExtreme lacks a comparable pre-period, and a small number
of uncertain records carry large sampling weights. Even so, the
cross-case evidence shows that destination activity, explicit migration
discourse, and broader governance response can diverge following
platform intervention; these dimensions should therefore be analyzed
separately.

% ============================================================
% REFERENCES
% ============================================================

\bibliographystyle{splncs04}
\bibliography{references}

\end{document}